\documentclass[sigconf]{acmart}

\usepackage{subfig, float}
\usepackage{amssymb}
\usepackage{amsmath}
\usepackage{graphicx}
\usepackage{array}
\usepackage{balance}
\usepackage{colortbl}
\usepackage{makecell}
\usepackage{bm}
\usepackage{wrapfig}

\usepackage[inline]{trackchanges}
\addeditor{DJORDJE}
\addeditor{BRANDON}
\addeditor{TIAN}
\addeditor{BHARAT}
\addeditor{ALAN}

\AtBeginDocument{%
  \providecommand\BibTeX{{%
    \normalfont B\kern-0.5em{\scshape i\kern-0.25em b}\kern-0.8em\TeX}}}

\copyrightyear{2020}
\acmYear{2020}
\setcopyright{acmcopyright}\acmConference[CIKM '20]{Proceedings of the 29th ACM International Conference on Information and Knowledge Management}{October 19--23, 2020}{Virtual Event, Ireland}
\acmBooktitle{Proceedings of the 29th ACM International Conference on Information and Knowledge Management (CIKM '20), October 19--23, 2020, Virtual Event, Ireland}
\acmPrice{15.00}
\acmDOI{10.1145/3340531.3412689}
\acmISBN{978-1-4503-6859-9/20/10}

\begin{document}

\title{Bid Shading in The Brave New World of First-Price Auctions}
\fancyhead{}





\author{Djordje Gligorijevic*, Tian Zhou*, Bharatbhushan Shetty, Brendan Kitts, Shengjun Pan, Junwei Pan and Aaron Flores}
    \thanks{*Authors contributed equally}
    \affiliation{ 
      \institution{Yahoo Research, Verizon Media}
      \city{Sunnyvale}
      \state{CA}
       \country{USA}
    }
    \email{{djordje.gligorijevic, tian.zhou, bharatbs,  brendan.kitts, alanpan, jwpan, aaron.flores}@verizonmedia.com}

\renewcommand{\shortauthors}{Gligorijevic, Dj. et al.}

\begin{abstract}
Online auctions play a central role in online advertising, and are one of the main reasons for the industry's scalability and growth. With great changes in how auctions are being organized, such as changing the second- to first-price auction type, advertisers and demand platforms are compelled to adapt to a new volatile environment. Bid shading is a known technique for preventing overpaying in auction systems that can help maintain the strategy equilibrium in first-price auctions, tackling one of its greatest drawbacks. In this study, we propose a machine learning approach of modeling optimal bid shading for non-censored online first-price ad auctions. We clearly motivate the approach and extensively evaluate it in both offline and online settings on a major demand side platform. The results demonstrate the superiority and robustness of the new approach as compared to the existing approaches across a range of performance metrics.
\end{abstract}




\maketitle

\section{Introduction}

Since the early days of the online advertising, the industry has faced the challenge of selling and distributing relevant ads to users at scale. The predominant solution has been to facilitate these efforts through online auctions organized by ad exchanges. The most widely used online auction format has been the generalized second-price auction (SPA), a powerful medium for advertisers to bid for reaching their target audiences with relevant product/service ads. These second-price auctions have become the engine for the online advertising industry, used for over 20 years, and driving worldwide advertiser revenues to over \$250 Billion in 2018 \cite{emarketer2018}, with more than \$100 billion in the US alone\footnote{\url{https://www.iab.com/wp-content/uploads/2019/05/Full-Year-2018-IAB-Internet-Advertising-Revenue-Report.pdf}, accessed January 2020}. 

In the second-price auction, when a user places a bid, they are charged the price of the next highest competing bidder plus (usually) a penny; so for example, if they bid \$5.00, and the next highest bidder is \$2.50, then they are charged \$2.51. SPAs exhibit the Vickrey property \cite{wilkens:cinderella}, which states that bidding ones true value is a dominant strategy - meaning that this will produce the best possible payoff regardless of actions taken by other bidders. In SPAs advertisers may simply focus
on calculating the value of the incoming impression and then bidding that true value.

However, midway through 2017, this situation suddenly changed. Many ad exchanges began switching to generalized first-price auctions (FPAs), a different bidding format that was very popular in the first years of online advertising~\cite{edelman2007internet}. In January 2017 there were no known FPAs used by major display ad exchanges \cite{kitts:fpa}, while between January and December 2019, the percent of auctions running on FPAs had risen from an astonishing 40\% \cite{benes:prices} to nearly 100\% \cite{google:rtb}. 
Several factors conspired to drive the industry towards the adoption of the FPA,
with the most important ones being increased demand for transparency and accountability \cite{chari1992us,sluis:guardiansuesrubicon,getintent:rtb1}.
In the first-price auction, there would be no possibility of an ad exchange improperly manipulating the clearing prices, as the price charged would always be exactly equal to the price offered. FPAs also allowed ad exchanges to capture all of the revenue from buyers since there would no longer be any discounting; this was an extremely attractive option for ad exchanges.

However, FPAs also carry some significant problems that were very notable in the early days of online advertising.
They introduce considerable complexity to bidding systems, in particular because they do not have the Vickrey property and have no strategy equilibrium (the easiest way to win an auction is to be fast and frequent in \textit{revising} bids with respect to the competitors bids). They are, thus, susceptible to system gaming strategies by different parties, which easily create volatile prices that in turn cause allocative inefficiencies \cite{edelman2007internet}.

Most importantly, in first-price auctions there is the fear of overpaying.
If an advertiser bids \$5.00 for an impression, and the next highest competing bid is \$2.50, they will get charged \$5.00 - but a strategic bid of \$3.00 would save the advertiser money. This means that the bidder has to introduce a whole new system - after calculating the value of the impression, the bidder then has to adjust their submitted bid downwards, so that if they win, they are charged only the minimum necessary to win their bid. Bidding too low though, increases the chances of not winning the auction at all, hence there is an intrinsic trade-off between the likelihood of winning and the payment in case it is won that needs to be taken into account. This practice of making a final adjustment to the bid price is referred to in the auction literature as \textit{Bid Shading}, and it is a frequent practice in many auction systems such as cattle auctions~\cite{zulehner2009bidding,pownall2013bidding}, US treasury auctions~\cite{hortaccsu2018bid} and FCC spectrum auctions~~\cite{chakravorti1995auctioning}.
The difference between the advertiser's private value and shaded bid for won auctions is called the \textit{bid surplus}, and optimization of this quantity is a major objective for the advertiser.



Online advertising auctions are dynamic, affected by a range of external factors, not to mention the arrival and departure of different bidders, bid shading is thus a very difficult problem.





In this study, we discuss a machine learning based bid shading approach for \textit{open (non-censored) first-price auctions} -- a type of auctions where there is a feedback containing minimum bid to win sent to all participants regardless of an auction outcome. 
Unlike closed FPAs where the price of the highest competing bidder is censored and SPAs where the price is known only for won auctions, in the \textit{open FPAs} it is always shared. 
Moreover, as in the early days of online advertising, \textit{open FPAs} are becoming a dominant type of first-price online auctions with the largest ad exchanges adopting it early.
The main benefit of this type of auctions is that they shed light on the bidding landscape and competition in such a manner that demand platforms can fairly and transparently optimize their surplus and submit appropriate bids. 

The approach this study discusses learns from historical non-censored auction data using features available at the ad opportunity, so as to estimate the \textit{optimal bid shading ratio} defined as ratio of the highest competing bid and calculated bid value. This approach is successfully deployed in a major Demand Side Platform (DSP), and we demonstrate its effectiveness in a real production environment, compared against more traditional approaches, and using a range of relevant metrics in offline and online settings.

Contribution of this study are enlisted below:
\begin{itemize}
    \item We propose and characterize the problem and challenges of estimating the optimal bid shading ratio for the open (non-censored) first-price online auctions.
    \item We propose efficient Factorization Machines based approach for optimizing the bidding surplus to model the historical auction data and estimate the shading ratio in real time. To the best of our knowledge this paper is the first to describe a working bid shading algorithm for bidding systems. 
    \item Improvements of total surplus show 18.57\% and 20.54\% for offline results and online results as compared to the existing models in production.
\end{itemize}

\section{Bid Shading Problem Definition, Recent Work and Challenges}
\label{sec:background}


\subsection{Canonical bidding optimization objective}
\label{sec:bidding}



In order to show ads to consumers, advertisers rely on DSPs (such as Google DoubleClick or Verizon Media DSP) to deploy highly efficient auction bidding infrastructures and run campaigns and lines on their behalf. Campaigns and lines can target an activity (such as click or conversion) and can optimize a range of objectives such are cost-per-X (view, click, mille, acquisition, install, etc.). Furthermore, each line is evaluated with key performance indicators (KPIs) which in addition to line objectives can measure total spend, win rate, etc.
Demand platforms are responsible for finding the best ad opportunities to deliver advertisers' ads, bidding for those opportunities, while spending as much as possible advertisers' budget to maximize delivery.




Online ad auction participants can submit a single bid in dollar value trying to win the opportunity to display an ad to a user. To achieve this, the following optimization objective (simplified for brevity) is formalized across all ad opportunities $i$:
\begin{equation} \label{eq:bidding}
\begin{aligned} 
    & \underset{b_i}{\arg\max} \sum_{i = 1}^{N} \mathbf{I} (b_i) * v_i \\
    & \text{subject to:} \sum_{i = 1}^{N} \mathbf{I} (b_i) * c_i \leq B,
\end{aligned}
\end{equation}
where $b_i$ is the bid value, $c_i$ is the payment to the exchange if the auction is won, indicated by $\mathbf{I} (b_i)$, and $v_i$ is the expected value to the advertiser of showing the intended ad for impression opportunity $i$. 
Typically, that value is defined in terms of an action associated to the ad impression, such as a click, view, conversion, etc.
For example, for conversion lines $v_i = pCVR_i *event\_value$, where $pCVR_i$ is the estimated probability of a conversion, and $event\_value$ is the monetary value of such conversion event. Finally, the constraint is placed on the optimization function to ensure that the total sum of impression costs to the advertiser $c_i$ does not exceed the budget $B$ defined for the same time window.



To solve the problem statement in (\ref{eq:bidding}), 
$b_i$ is often calculated as:
\begin{equation}
	b_i = \alpha * v_i,
\end{equation}
where $\alpha$ is summarizing multiplicative parameter that control the delivery of a line, i.e. with respect to the bid shading, delivery pacing and KPIs. Depending on the auction type, $b_i$ can be unshaded (annotated with $b_i^u$) or shaded.
Bid shading, thus, plays a key role in deciding on the bid value as a part of multiplicative term $\alpha$. 

\subsection{Bid Shading Maximization}
\label{sec:bid_shading}

For first-price auctions, the optimal bidding strategy $f$ attempts to  decrease (shade) the bid price as much as possible to the minimum bid necessary to win. This process is called bid shading.

The optimization function of the bid shading can be defined through maximizing the surplus:
\begin{equation} \label{eq:surplus_per_response}
\text{bid surplus} = \sum_{i=1}^{N}(b_i^u - b_i)\mathbf{I}(b_i),
\end{equation}
where bid $b_i^u$ is the unshaded bid.
It should be noted that this formulation of the problem implies that bid shading is the process of decreasing the bid value only, as increasing the bid value would negatively coact with the process of estimating the value of an impression.


\subsection{Prior bid shading algorithms}

\subsubsection{Supply-side bid shading algorithms}

The shift from second to first-price auctions was extremely rapid, and many advertisers may have difficulty calculating shading adjustments. To ease the transition to first-price, a range of ad exchanges have begun to offer "default bid shading services" to help advertisers who might struggle to implement their own algorithm. Examples include Google's ADX ``Bid Translation Service''~\cite{google:rtb}, Rubicon's ``Estimated Market Rate''~\cite{rubicon:EMR}, and AppNexus's ``Bid Price Optimization'' system ~\cite{appnexus:BPO}. These services are designed to shade high, private-value-like bids down to a price at which the advertiser continues to win at an equivalent rate, but with a less egregious price. While this is a useful outcome for supply-side platforms, since it ensures impressions continue to sell, whilst avoiding what might be otherwise be considered ``price gouging'', it is most certainly not utility maximizing for advertisers. These services are only sometimes available, and discontinuation of these services is imminent once demand platforms' bid shading efforts have become sufficiently mature.


\subsubsection{Segmented, non-linear shading algorithm}
\label{sec:non_linear_bid_shading}

One approach that is simple but effective, is to estimate the optimal bid shading ratio using parametric functions of the unshaded bid. The function parameters are either tuned manually, or algorithmically, based on a feedback mechanism that tries to maximize surplus. Parameters are trained for each unique inventory segment, which is represented as a combination of salient inventory properties such as seller exchange, top level domain, and so on. For each of these inventory segments, historical surplus data is fed back to an algorithm that adjusts parameter values iteratively, attempting to maximize surplus for that segment. 
As a benchmark for the new methodology proposed in this paper, the following parametric function used in production by the DSP system was considered for both offline and online experiments:
\begin{equation}
    b_i=f(b_i^u) = \begin{cases}
            \log \frac{1 + u_1 * u_2 * b_i^u}{u_2} & u_2 > 0\\ 
            b_1 * b_i^u & \text{otherwise}
        \end{cases}
\end{equation}
where $u_1, u_2, b_1$ are the parameters, and $b_i^u$ is the unshaded bid. The algorithm used to adjust the parameters was a recursive least square methodology and segments were defined as the combination of exchange, top level domain, device and layout.

One drawback of the above segment-based, nonlinear approach, is that segments have to be predetermined and finding a suitable segment definition requires substantial analysis without being able to provide optimality guarantees.  Another disadvantage is that information across segments is not shared, which is a problem for segments that do not have enough traffic. 

\subsubsection{Related work on bidding landscape prediction}
Several approaches were developed for modeling censored data of SPAs to estimate the bidding landscape~\cite{wu2015predicting,wu2018deep}. These approaches have a goal to directly estimate minimum bid to win, using their bid price as lower and upper bound when auction is lost and won, respectively, through distributional assumptions (such as Normal or Gamma) on bid shading landscape.
The distributional assumption can be a very limiting factor in predicting bid landscape as we show in the following section, and the studies have not shown that any distribution tested is distinctly the best. Moreover, as these approaches are developed for the SPAs it would require extending them to the use case of FPA in order to conduct comparisons.




\subsection{Modeling challenges}
\label{sec:challenges}
For traditional machine learning problems such as CTR or CVR predictions a predictor is trained on established training data, and the assumption is that available features provide sufficient richness to describe the problem. Estimating the optimal bid shading factor for an ad opportunity, on the other hand, relies on not only drawing insights from historical auction data, but also foreseeing how the other bidders would \textit{play the game}. The following challenges are often barriers when we formulate and solve bid shading problems:
\begin{itemize}
    \item \textit{A large variability of the highest competing bid price}. Knowing the highest competing bid prices as provided by open bid auctions when bidding for their inventory is crucial piece of information to know the optimal bid ratio. However, the distribution of the highest competing bid prices can vary significantly across different sets as shown in Figure~\ref{fig:mb2w_examples}. Each graph in Figure~\ref{fig:mb2w_examples} shows the empirical distribution of the highest competing bid price (per mille) for a set of fixed page and user attributes, and the differences among different sets can provide insight into the variability of the problem bid shading algorithm needs to model.
        \begin{figure} [t!]
        \centering
        \subfloat{\includegraphics[width=0.25\textwidth]{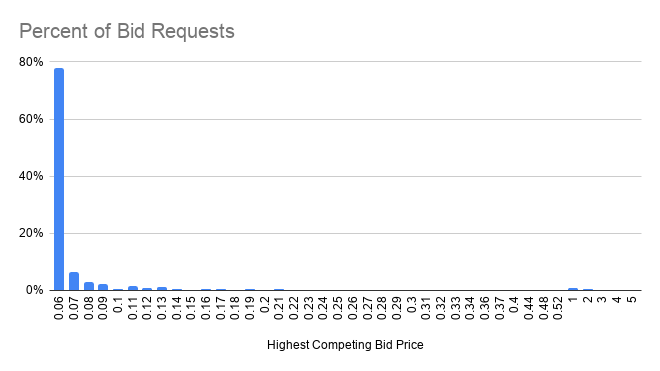}}
        \subfloat{\includegraphics[width=0.25\textwidth]{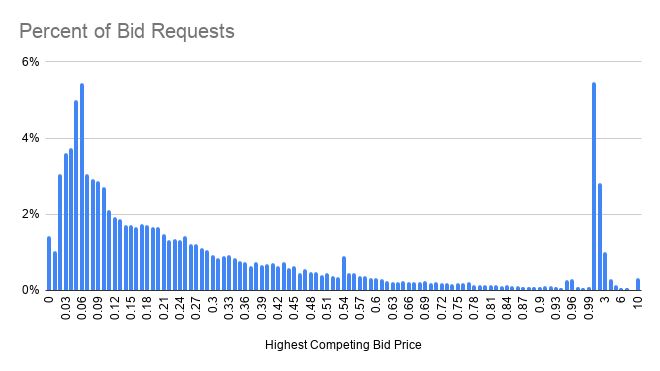}}\\
        \subfloat{\includegraphics[width=0.25\textwidth]{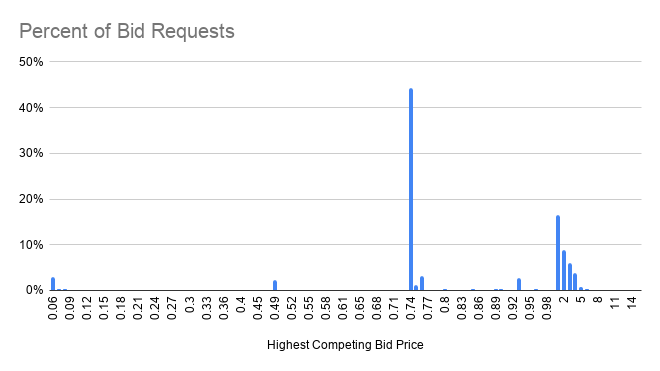}}
        \subfloat{\includegraphics[width=0.25\textwidth]{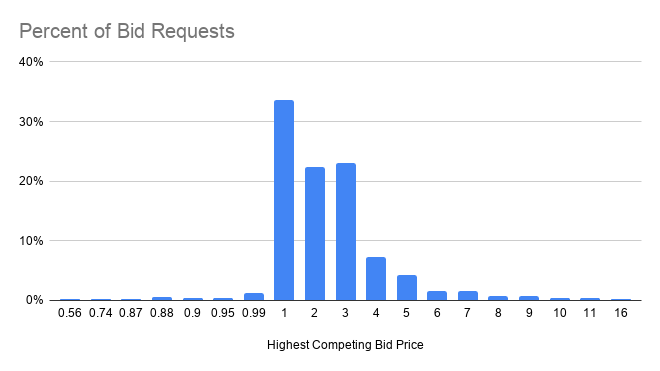}} \\
        \caption{Diversity of distributions of the highest competing bid prices for four fixed sets of Web page and user attributes.}
        \label{fig:mb2w_examples}
    \end{figure}

    \item \textit{Data collection}. In order to learn a bid shading strategy, it is important to log not only won but also lost bids information which increases storage requirements exponentially.
    \item \textit{One can never step into the same auction twice}. Every auction is unique. Machine learning approaches often make assumption that the test data will be generated from the same distribution that generated the training data. However, in an online ad auction system, ad inventory, end user, budget, and participating bidders are constantly changing, which poses challenges for any machine-learning approach. 
    \item \textit{Many other factors contribute to the final bid price}. Bid shading is only one step of many in a complex DSP system. For example, targeting efforts, CTR/CVR predictions, pacing control, etc., attempt to make adjustments to the final bid price. A bid shading algorithm is only useful if it's resilient to noises and changes in other parts of the DSP system.  
\end{itemize}

\section{emand platform Bid shading system overview}
\label{sec:system_overview}
\begin{figure}[ht!]
	\centering
	\includegraphics[width=0.48\textwidth]{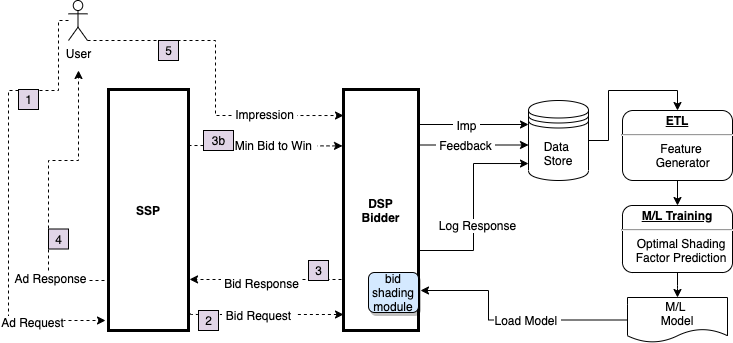}
	\caption{Overview of the bid shading in the bid generation system.}
	\label{fig:system_overview}
	\vspace{-15pt}
\end{figure}

We provide a system overview of the Verizon Media DSP which is one of the major DSPs in the US and the World with more than 200 billion daily requests received from more than 40 different ad exchanges. The DSP's production ad serving system workflow (of which the \textit{bid shading module} is part of) is shown in Figure~\ref{fig:system_overview}. On the left side of the DSP bidder we can see the sequence of events that occur for each ad opportunity generated by a user. User generates visits a web page, Supply Side Platform (SSP) generates ad request and sends it to several DSP's, the DSP will provide a first-price bid and it will receive information of minimum bid to win regardless of the bidding outcome. If bid was won, ad impression is made and logged in the data store together with the ad exchange feedback. On the backend side of the demand platform, a feature generating process is run with training machine learning model to predict optimal shading factor for different impressions once a day. The model is stored into a model file loaded by host machines in the bid shading module of the DSP bidder to be used in the following bid requests made by different SSP's. 

\paragraph{Requirements and limitations for predictive approaches}
This particular system setup allows for the models to be trained on very large datasets and stored in formats optimized for efficient serving in a very limited bid request response time frame.
During the total allowable 20ms response time, multiple modules needed to go through one by one including targeting, click/conversion prediction, internal auction, value capping, bid shading and others. The current programmatic bidding system in Verizon Media does not support parallel processing of these modules, which results in strict latency constraints of each module that we take into consideration during machine learning modeling phase.


\section{Predicting the Optimal Shading Ratio}
\label{sec:methodology}

\subsection{Methodologies}
In this section we describe the Factorization Machines (FM) model whose objective is to estimate the optimal bid shading ratio defined as the bid price divided by the minimum bid to win on historical bidding data. 
We then provide details of the loss function used designed to ensure surplus maximization.

\subsubsection{Factorization Machines}
Factorization Machines are a popular and efficient model used in recommendation systems. They have been very successfully used in applications involving sparse data such as CTR~\cite{pan2018field,juan2016field} and CVR~\cite{pan2019predicting} estimation outperforming popular second-degree polynomial models, while they maintain a terrific property of having only quadratic floating-operations complexity $O(N^2)$ in the space of data fields ($N$). With their high efficiency and effectiveness they are the approach that performs while complying with strict deployment restrictions discussed in Section~\ref{sec:system_overview}.

FMs learn a weight $w_i$ and an embedding vector $\bm{v}_i \in \mathbb{R}^K$ for each feature to model both its main effect and its interaction effects with other features, respectively. 
The interaction between two features $i$ and $j$ is modeled as the dot product $\langle \bm{v}_i, \bm{v}_j \rangle$ of their corresponding embedding vectors $\bm{v}_i$, $\bm{v}_j$. The formula is shown in Eq.~\ref{eq:fm}:

\begin{equation}
    \Phi_{FM}\left((\bm{w},\bm{v}),\bm{x} \right)=  w_0 + \sum_{i=1}^m x_i w_i +\sum_{i=1}^{m}\sum_{j=i+1}^m x_{i}x_{j} \langle \bm{v}_{i}, \bm{v}_{j}\rangle
    \label{eq:fm}
\end{equation}
Here, $K$ is the dimension of the embedding vectors, and it's usually a small integer, in our experiments $10$. 

The reason behind FMs success is that it can always learn some meaningful embedding vector for each feature as long as the feature itself appears enough times in the data, which makes the dot product a good estimate of the interaction effect of two features even if they have never or seldom occurred together in the data.

\subsubsection{Loss function and optimization}
Provided that the model's objective is to estimate the optimal bid shading ratio as its target $y$, a natural choice for a loss function was mean squared error loss:
\begin{equation}
    \mathcal{L} = \frac{1}{N}\sum_{i=1}^{N} \left( y_i - \Phi_{FM}(\bm{w},\bm{v}),\bm{x}_i) \right)^2
\end{equation}

However, the goal of the algorithm like this would be only to match the optimal bid shading ratio, whereas our goal is also to \textit{maximize the bidding surplus}, \textit{spend more}, 
and \textit{improve win rate}.
To that end we designed an asymmetric loss function \cite{tian2009forecasting} that:
\begin{itemize}
    \item Treats win/loss differently by penalizing more when we overshade and loose the bid, thus optimizing win rate and spend
    \item Has different loss weights for different bid requests such as penalizing less the higher bid surplus accumulation is achieved, thus optimizing the bidding surplus
    \item Has a ready capping hyperparameter that allows control over the penalty -- the more penalty when losing the bids will result in overall higher shading factor, higher win rate, and more spend. And vice versa.
\end{itemize}

We extend the simple quadratic loss function by imposing asymmetry between over-prediction and under-prediction as:
\begin{equation}
    \mathcal{L} = \frac{1}{N} \sum_{i=1}^{N} \left[(y_i - \Phi_{FM}(\bm{w},\bm{v}),\bm{x}_i))^2 * |\mathbb{I}_{\Phi_i < y_i} + \alpha| \right],
\end{equation}
where
\begin{equation}
    \mathbb{I}_{x} = \left\{\begin{matrix} 1 & \text{if } x \geq 0\\   -1 & \text{if } x < 0 \end{matrix}\right. ,
\end{equation}
and the optimization function becomes:
\begin{equation}
    \mathcal{L} = \frac{1}{N}\sum_{i=1}^{N} \left[(y_i - \Phi_{FM}(\bm{w},\bm{v}),\bm{x}_i))^2 * \left|\left\{\begin{matrix} 1 + \alpha & \text{if loss of the bid} \\  1 - \alpha & \text{if win of the bid} \\  \end{matrix}\right. \right| \right].
\end{equation}

The parameter $\alpha \in (0,1)$ indicates the degree of asymmetry. For instance, if $\alpha > 0$ the model suffers more loss from under-prediction than from over-prediction. The further $\alpha$ is from 0, the more asymmetric the model's preference between under-prediction and over-prediction is. In our definition, $\alpha$ is obtained as:
\begin{equation}
    \alpha = \text{min} (1, \text{max} (\text{opt\_surplus}, \gamma)), \gamma \in (0, 1),
\end{equation}
where
\begin{equation}
    \text{opt\_surplus = price before shading - minimum price to win}
\end{equation}
is normalized to [0, 1] scale to match the range of $\alpha$.
The hyperparameter $\gamma$ used for flooring $\alpha$ by allowing a small manual control over the shading factor predictions in the production. We thus retain control over some performance metrics not included in the loss function above to preserve its generalization across different demand platform optimization goals.

Parameters $(\bm{w},\bm{v})$ can be easily trained end-to-end through standard gradient descent approaches as the proposed asymmetric loss function preserves the smoothness of the MSE loss.

\section{Experimental evaluation}
\label{sec:experiments}
A detailed analysis in both offline and online setting of the proposed methodology as compared against several baselines across different metrics is provided here.

All algorithms have been trained on a sample of 7 days of data and evaluated on the next day using won bids only. The datasets sample approximately has 9 million auction feedback per day. We are not using the feedback from auctions that were lost, because in the current bidding system definition (Section~\ref{sec:bidding}), the bid shading function is supposed to optimize only the bid surplus given the bid value determined through other modules, thus only lowering the bid price is allowed (i.e. we do not wish to pay for opportunity more that it is worth). 

For input features algorithms use as inputs to predict the optimal shading ratio for each ad opportunity we are using 13 publisher and context fields available during the ad call:
\begin{itemize}
    \item Publisher fields: $Page\_TLD$, $Subdomain$, $Publisher\_ID$ and $Request\_publisher\_ID$,
    \item Context fields: $Country\_ID$, $Day\_of\_Week$, $Hour\_of\_Day$, $Device\_type\_ID$, $App\_name$, $Is\_new\_user$, $Target\_deal\_ID$, $Layout\_ID$ and $Ad\_position\_ID$.
\end{itemize}

These 13 fields amount to hundreds of thousands of total categorical features used by the algorithms.


\subsection{Baselines}
For the machine learning approaches, we considered the following: 1) Linear regression (LR), 2) Multi Layer Perceptron regression (MLP) and 3) Factorization Machines regression (FM). All algorithms are optimized using the aforedescribed asymmetric loss function. 

\subsection{Metrics}
Below we describe a spectrum of metrics used to evaluate performance of the bid shading algorithms from perspectives of regression performance and demand platform performance metrics.

\subsubsection{Regression performance metrics}
Selected metrics for evaluating regression performance of proposed algorithms in this study are:
\begin{itemize}
\item Mean Squared Error (MSE): 
\begin{equation}
 MSE = \frac{1}{N}\sum_{i=1}^{n}(y_i - \hat{y}_i)^2   
\end{equation}
\item Coefficient of determination ($r^2$): 
\begin{equation}
 r^2 = \frac{n\Sigma_{xy} - \Sigma_x \Sigma_y}{\sqrt{(n\Sigma_{x^2} - (\Sigma_x)^2)(n\Sigma_{y^2} - (\Sigma_y)^2)}}   
\end{equation}
\end{itemize}

\subsubsection{Demand platform performance metrics}
The proposed regression task metrics do not encapsulate the most important aspects of the performance of the bid shading system. As discussed in Section~\ref{sec:bid_shading}, the bid shading is actually a process of optimizing the eq.~\ref{eq:surplus_per_response}. To that end we propose the following metrics of evaluation (using the consistent notation) that we will consider the main performance metrics:
\begin{itemize}
    \item Total surplus per bid response: 
    \begin{equation}
        surplus = \sum_{i=1}^{N}(b_i^u - b_i)\mathbf{I}_{B^*}(b_i)
    \end{equation}
    \item Total spend: 
    \begin{equation}
        total\_spend = \sum_{i=1}^{N}b_i\mathbf{I}_{B^*}(b_i)
    \end{equation}
    \item Win rate: 
    \begin{equation}
        win\_rate = \frac{1}{N}\sum_{i=1}^{N}\mathbf{I}_{B^*}(b_i)
    \end{equation}
    \item Cost per mille (CPM): 
    \begin{equation}
        CPM = \frac{1}{N}\sum_{i=1}^{N}b_i\mathbf{I}_{B^*}(b_i)
    \end{equation}
\end{itemize}
Results reported on these metrics are percentage improvements or increases as compared to production's non linear bid shading algorithm (Section~\ref{sec:non_linear_bid_shading}) as absolute numbers could not be disclosed.

\subsection{Offline experiments and analysis}
In this section, we will analyze the performance of the machine learning based approaches using both regression metrics and demand platform performance metrics in an offline setting using the retrospective analysis. This will allow us to very easily characterize algorithmic performance on the available data, as well as to do a deeper dive into goal type performance analysis and fine-tuning of hyperparameters.

\subsubsection{Machine learning based approaches and their performance}
On the described dataset we train the three mentioned machine learning based approaches, characterize their regression performance and finally compare their individual improvements against the production baseline.
The regression performance is aimed to evaluate how well the algorithms fit the data, which, in the setup proposed in this study, means predicting the competitive landscape, a very difficult task indeed given a large number of demand platforms and highly variable number of external lines and ads that could potentially target the current opportunity. This is clearly reflected in lower $r^2$ values on all algorithms, however its positivity shows that models can still capture some of the large variability of the bid shading signal.

\begin{wrapfigure}{r}{0.30\textwidth}
  \begin{center}
    \includegraphics[width=0.30\textwidth]{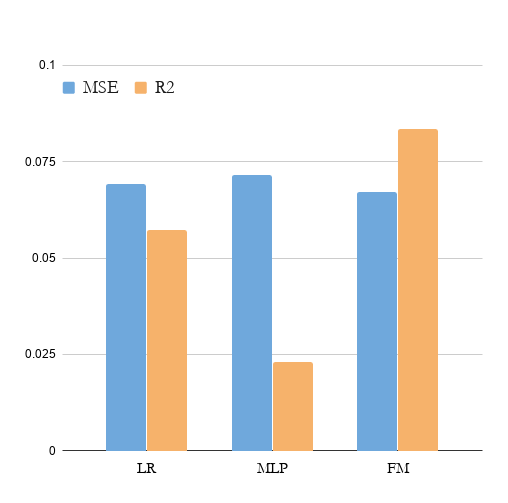}
  \end{center}
  \caption{MSE and $r^2$ performance metrics for the three machine learning based approaches.}
  \label{fig:exp_main_regression}
  \vspace{-15pt}
\end{wrapfigure}
For the regression metrics (Figure~\ref{fig:exp_main_regression}) we can see that there is a distinction among the algorithms (especially in terms of the $r^2$ metric) where the MLP approach had the worst performance, while the proposed FM approach had the best performance. We suspect that the combination of existing features does not provide as much as predictive information as does modeling heterogeneity of each feature (i.e. through the feature learning layer in the FM model).



More interesting results are shown in Table~\ref{tbl:exp_main_performance}, where the three machine learning based approaches are directly compared against the non-linear bid shading used in production on surplus, spend, win rate and CPM metrics. Across the board we can see that each machine learning based approach outperforms the nonlinear bid shader on surplus, total spend and win rate. 
The CPM metric is naturally increased with the increase of spend and win rate and in production environment its value is controlled in several goal type lines such as eCPM which we will show in our online experiments.  Moreover we additionally control the balance between spend and CPM increase through hyperparameters of the algorithms (in particular the $\gamma$ parameter) which we discuss in Section~\ref{sec:gamma_tuning}.
\begin{table}[h!]
\resizebox{0.48\textwidth}{!}{%
\begin{tabular}{lcccc}
\toprule
& Surplus imp. & Spend imp. & CPM inc. & Win rate impr.       \\
\midrule
LR           & 17.20\%            & 128.63\%   & \textbf{67.56\%}   & \textbf{36.40\%} \\
MLP          & 16.18\%            & \textbf{136.67\%}   & 73.85\%   & 36.12\% \\
FM           & \textbf{18.57\%}            & \textit{133.02\%}   & \textit{70.82\%}   & \textbf{36.40\%} \\
\bottomrule
\end{tabular}}%
\caption{Improvements across surplus, total spend, win rate and increase of the CPM expressed in percentages over the nonlinear bid shading production baseline as obtained by the three machine learning based approaches.}
\label{tbl:exp_main_performance}
\vspace{-10pt}
\end{table}
The highest surplus and win rate improvements are obtained with the factorization machine approach yielding second highest spend at a bit higher CPM increase than of the linear regression. 

\subsubsection{Performance analysis per demand platforms' goal type}
Further characterization of the performance of different models can be done per goal types lines are optimizing towards, primarily because of the two following reasons: 1) the bid price calculation is slightly different for each goal type and 2) each goal type can have different control signals. 
The goal types of particular interest on the platform with their prevalence on the test set are \textit{None} goal type with 40.5\% and five CP-X goals -- CPC (click) with 12.12\%, CPA (action) with 38.95\%, CPCV (completed view) with 0.57\%, CPViewI (viewed impressions) with 0.78\% and eCPM (mille) with 3.60\%. 

\begin{figure}[ht!]
	\centering
	\includegraphics[width=0.5\textwidth]{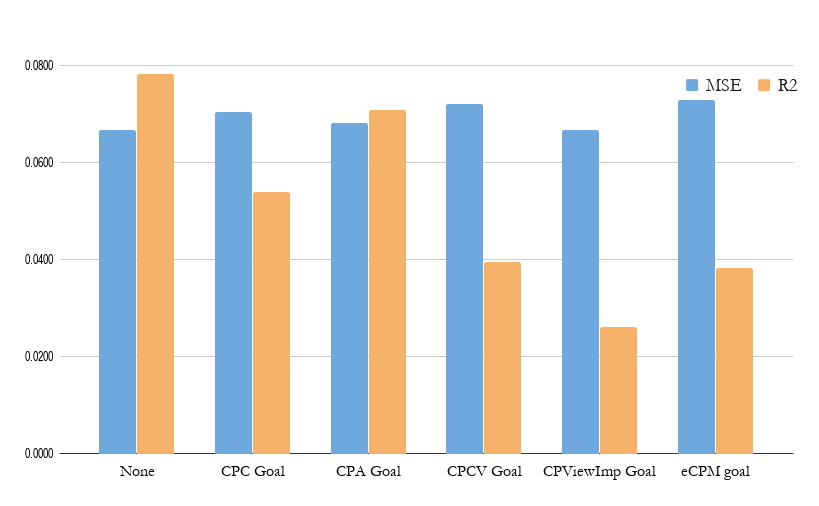}
	\caption{MSE and $r^2$ performance metrics for the Factorization Machine algorithm broken down by the goal type.}
	\label{fig:exp_fm_goal_type_regression2}
	\vspace{-10pt}
\end{figure}

Again, we separate the evaluation of algorithms into regression and platform performance metrics and show the performance of the proposed FM model as the best performing model. Regression metrics per goal type are shown in Fig.~\ref{fig:exp_fm_goal_type_regression2}. We can see that regression performance does vary per line type, where the performance is most aligned with the overall results for two dominant goal types -- None and CPA. It is expected for the performance numbers to drop for goal types that have a very small yet non-negligible share of the serving opportunities due to the inherent noise of the problem.

Furthermore, once we analyze the performance of the FM-based bid shading approach against the nonlinear production baseline across surplus, total spend, win rate and CPM we can clearly see that there are improvements on the key metrics.
\begin{figure}[h!]
	\centering
	\includegraphics[width=0.5\textwidth]{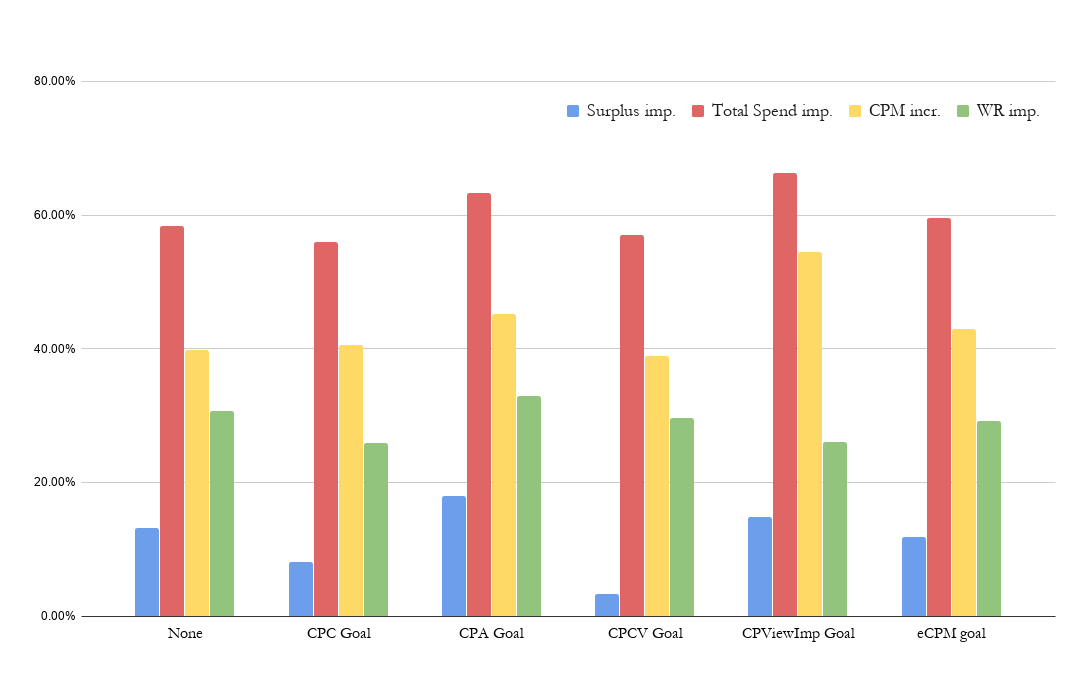}
	\caption{Improvements across surplus, total spend, win rate and increase of the CPM expressed in percentages over Nonlinear bid shading production baselines obtained by the Factorization Machine model broken down by the goal type.}
	\label{fig:exp_fm_goal_type_performance2}
\end{figure}
Moreover, as the highest increase in surplus we observe is for the CPA goal which typically has the highest prices, we can concur that better shading can deliver gains in practice for advertisers targeting sophisticated, rare, and high-priced goals. But also for goal types with smaller values such as eCPM, or even CPC, where shading has fewer degrees of freedom to pick and chose how to adjust the price we can clearly see that there is added value.
As a result of this analysis, we can conclude that the FM model provides a clear advantage over the production model across different goal types.


\subsubsection{Using parameter $\gamma$ to control the performance metrics}
\label{sec:gamma_tuning}
For successful deployment in production environment we allowed a control over the algorithm through parameter $\gamma$ acting as a floor to asymmetric parameter $\alpha$. By changing $\gamma$ we can have a control over models performance, i.e. choosing one with the best bid surplus or lowest CPM. This parameter is meant to select the model parameters that will be aligned with the control signals of production system.
For the FM algorithm, we searched for the best $\gamma$ parameter by increments of $0.2$ (results shown in Table~\ref{tbl:gamma_impact}).
\begin{table}[h!]
\resizebox{0.48\textwidth}{!}{%
\begin{tabular}{lllllll}
\toprule
$\gamma$ value    & 0      & 0.2    & 0.4    & 0.6    & 0.8    & 1      \\
\midrule
Surplus impr. & 17.77\%  & \textbf{18.57\%}  & 17.51\%  & 17.31\%  & 17.20\%  & 17.49\%  \\
Spend impr.   & 144.02\% & 133.02\% & 146.57\% & 148.27\% & \textbf{149.30\%} & 147.28\% \\
CPM inc.              & 75.24\%  & \textbf{70.82\%}  & 75.65\%  & 76.01\%  & 76.25\%  & 75.83\%  \\
Win rate impr.      & 39.25\%  & 36.40\%  & 40.36\%  & 41.03\%  & \textbf{41.42\%}  & 40.61\% \\
\bottomrule
\end{tabular}}%
\caption{Impact of $\gamma$ parameter evaluated across 6 values on 0.2 increment for the Factorization Machine algorithm.}
\label{tbl:gamma_impact}
\vspace{-15pt}
\end{table}
In this experiment we obtained the highest surplus improvement and the lowest CPM increase at $\gamma = 0.2$ operating point, while the highest total spend and win rate are obtained at $\gamma = 0.8$ operating point, at the cost of the highest CPM increase and lowest surplus improvement. With respect to the criteria described above, we selected $\gamma = 0.2$ as a parameter for all algorithms throughout the experiments.

Using the asymmetric loss function with $\gamma = 0.2$ compared to symmetric mean square error yielded an increase of 0.1\% and decrease of 1.2\% for MSE and $r^2$, respectively, while it provided a increase of 6\% for total surplus, and a 6.6\% decrease of CPM (with <10\% decrease in win rate and spend suggesting that this approach focuses more on higher value bids).


\subsection{Online experiments and analysis}
Offline experiments were able to show retroactive performance of the proposed bid shading approach while lacking any interaction with the other modules of the system. 
Finally, we discuss performance of the proposed bid shading strategy in the online setting. The two LR and two FM models were deployed on distinct servers (we removed MLP model due to its high computational cost and poor performance in offline experiments), serving independent traffic and performance was monitored over several days.


\subsubsection{Online A/B test}
\begin{figure}[t!]
	\centering
	\includegraphics[width=0.43\textwidth]{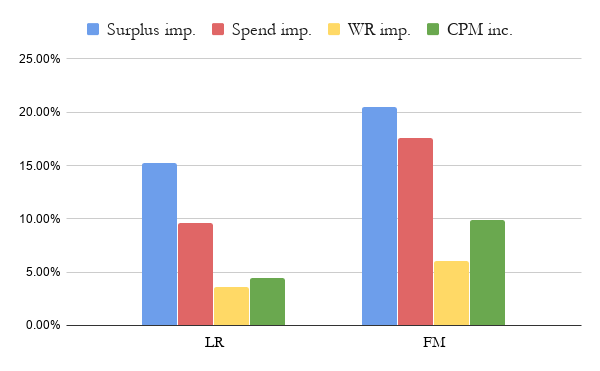}
	\caption{Performance metrics improvements and increases (expressed in percentages) for LR and FM model as compared to the nonlinear bid shared collected from separate serving buckets (A/B test).}
	\label{fig:exp_online_ab}
	\vspace{-15pt}
\end{figure}

In terms of online AB tests, we aggregated results for both LR and FM models and compared them to the randomly selected traffic volume of same scale. Results obtained by the two models as compared to non-linear bid shading approach are given in Figure~\ref{fig:exp_online_ab}. Online results are similar to the offline results in terms of relative ranking of models, with the proposed FM approach being the best one. However, thanks to the online serving control the online setting provided similar or better surplus improvements but at a much lower CPM increase ($< 10\%$ compared to $> 70\%$ in the offline counterparts). This behavior is very desirable as the machine learning based bid shading module is capable of playing well into the system providing desired improvements on key metrics.

\subsubsection{Online A/A/B/B tests}
\begin{figure}[h!]
	\centering
	\includegraphics[width=0.46\textwidth]{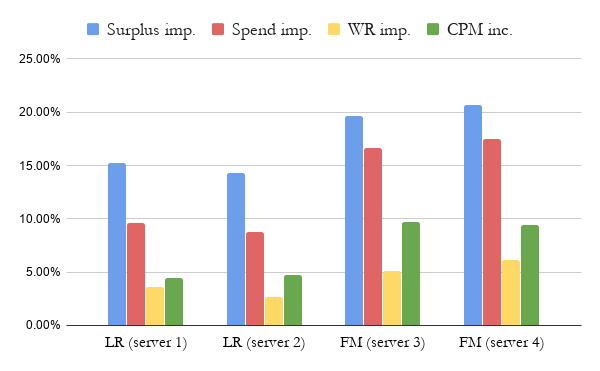}
	\caption{Performance metrics improvements and increases (expressed in percentages) for LR and FM model as compared to the nonlinear bid shading collected from two different servers each showing stability of performance of ML-based approach (A/A/B/B test).}
	\label{fig:exp_online_aa}
	\vspace{-15pt}
\end{figure}

To test the stability, we separately aggregated performance of the two algorithms across different servers and thus created an A/A/B/B test. In Fig.~\ref{fig:exp_online_aa} we can see that the performance increases compared to the non linear bid shader is very similar between servers running the same algorithm, be it LR or FM model. We can thus strengthen the claim that the performance measured of any of the algorithms was not a fluke.

\begin{figure}[h!]
	\centering
	\subfloat{\label{fig:assym_regression}\includegraphics[width=80mm]{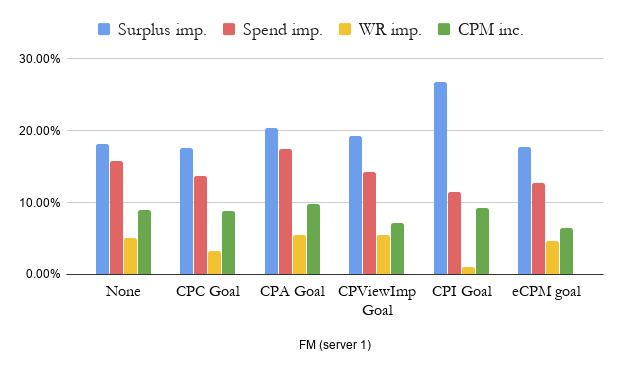}}\hspace{2mm}
	\vspace{-10pt}
	\subfloat{\label{fig:assym_performance}\includegraphics[width=80mm]{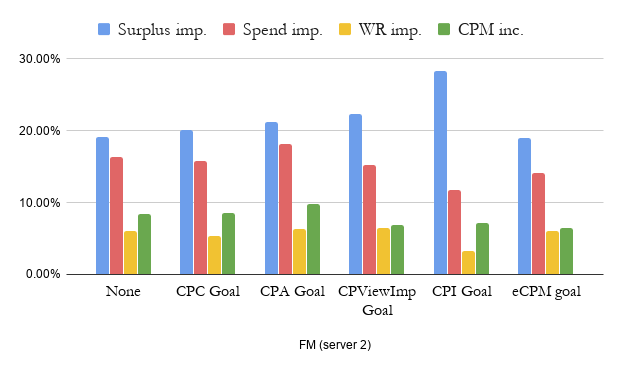}}
	\caption{Performance metrics improvements and increases (expressed in percentages) for the FM model across 6 available goal types as compared to the nonlinear bid shared collected from three separate serving buckets (A/A/B/B test).}
	\label{fig:exp_online_goal_type_aab}
	\vspace{-10pt}
\end{figure}

Finally we ran a similar setup of experiments only per goal type (similarly to the offline counterpart) with sufficient traffic available for the best performing FM model. Figure~\ref{fig:exp_online_goal_type_aab} shows that across different goal types and across two different servers, the FM model provides very stable and high increase in bid surplus, total spend and win rate, while maintaining relatively low CPM increase ($<10\%$).
With the last online A/A/B/B test we conclude that the machine learning based bid shading approach using the FM model is superior to any other approach considered for the demand platform's bidding system, and that it provides a significant value of allowing the strategy equilibrium in the non-censored first-price auctions.

\section{Conclusions and future directions}
\label{sec:conclusions}
In this study, we focused on describing the problem of estimating optimal shading factor using machine learning models designed for the non-censored type of first-price auctions, which are becoming the widespread type for online advertising. The bid shading problem under such circumstances becomes a point estimate problem with many opportunities and challenges. We described an approach that fitted well within DSP's production environment and performed exceptionally. The proposed FM approach with asymmetric loss function provided valuable improvements over other approaches both non-linear shading and machine learning based in both offline and online settings. The proposed approach brought improvements in estimated surplus of $18.57\%$ and $20.54\%$ in offline and online setting respectively, while providing stability of performance per goal type and also on A/A tests in an online setting. 
However many challenges still remain: Different goal types weight metrics differently which could be taken into account when modeling; furthermore, using information from bids lost
could be used to improve total spend and performance for lines by modeling the bidding competition.



\balance
\bibliographystyle{ACM-Reference-Format}
\bibliography{references}


\begin{thebibliography}{21}


\ifx \showCODEN    \undefined \def \showCODEN     #1{\unskip}     \fi
\ifx \showDOI      \undefined \def \showDOI       #1{#1}\fi
\ifx \showISBNx    \undefined \def \showISBNx     #1{\unskip}     \fi
\ifx \showISBNxiii \undefined \def \showISBNxiii  #1{\unskip}     \fi
\ifx \showISSN     \undefined \def \showISSN      #1{\unskip}     \fi
\ifx \showLCCN     \undefined \def \showLCCN      #1{\unskip}     \fi
\ifx \shownote     \undefined \def \shownote      #1{#1}          \fi
\ifx \showarticletitle \undefined \def \showarticletitle #1{#1}   \fi
\ifx \showURL      \undefined \def \showURL       {\relax}        \fi
\providecommand\bibfield[2]{#2}
\providecommand\bibinfo[2]{#2}
\providecommand\natexlab[1]{#1}
\providecommand\showeprint[2][]{arXiv:#2}

\bibitem[\protect\citeauthoryear{AppNexus}{AppNexus}{[n.d.]}]%
        {appnexus:BPO}
\bibfield{author}{\bibinfo{person}{AppNexus}.}
  \bibinfo{year}{[n.d.]}\natexlab{}.
\newblock \showarticletitle{Demystifying Auction Dynamics for Digital Buyers
  and Sellers}.
\newblock


\bibitem[\protect\citeauthoryear{Benes}{Benes}{2018}]%
        {benes:prices}
\bibfield{author}{\bibinfo{person}{R. Benes}.} \bibinfo{year}{2018}\natexlab{}.
\newblock \showarticletitle{First-Price Auctions Are Driving Up Ad Prices: Ad
  buyers should adjust their bidding strategies}.
\newblock \bibinfo{journal}{\emph{eMarketer}}.
\newblock
\urldef\tempurl%
\url{https://www.emarketer.com/content/first-price-auctions-are-driving-up-ad-prices}
\showURL{%
\tempurl}


\bibitem[\protect\citeauthoryear{Chakravorti, Sharkey, Spiegel, and
  Wilkie}{Chakravorti et~al\mbox{.}}{1995}]%
        {chakravorti1995auctioning}
\bibfield{author}{\bibinfo{person}{Bhaskar Chakravorti},
  \bibinfo{person}{William~W Sharkey}, \bibinfo{person}{Yossef Spiegel}, {and}
  \bibinfo{person}{Simon Wilkie}.} \bibinfo{year}{1995}\natexlab{}.
\newblock \showarticletitle{Auctioning the airwaves: the contest for broadband
  PCS spectrum}.
\newblock \bibinfo{journal}{\emph{Journal of Economics \& Management Strategy}}
  \bibinfo{volume}{4}, \bibinfo{number}{2} (\bibinfo{year}{1995}),
  \bibinfo{pages}{345--373}.
\newblock


\bibitem[\protect\citeauthoryear{Chari and Weber}{Chari and Weber}{1992}]%
        {chari1992us}
\bibfield{author}{\bibinfo{person}{V Chari} {and} \bibinfo{person}{Robert
  Weber}.} \bibinfo{year}{1992}\natexlab{}.
\newblock \showarticletitle{How the US Treasury should auction its debt}.
\newblock \bibinfo{journal}{\emph{Federal Reserve Bank of Minneapolis Quarterly
  Review}} \bibinfo{volume}{16}, \bibinfo{number}{4} (\bibinfo{year}{1992}).
\newblock
\urldef\tempurl%
\url{http://kylewoodward.com/blog-data/pdfs/references/chari+weber-quarterly-review-1992A.pdf}
\showURL{%
\tempurl}


\bibitem[\protect\citeauthoryear{Edelman, Ostrovsky, and Schwarz}{Edelman
  et~al\mbox{.}}{2007}]%
        {edelman2007internet}
\bibfield{author}{\bibinfo{person}{Benjamin Edelman}, \bibinfo{person}{Michael
  Ostrovsky}, {and} \bibinfo{person}{Michael Schwarz}.}
  \bibinfo{year}{2007}\natexlab{}.
\newblock \showarticletitle{Internet advertising and the generalized
  second-price auction: Selling billions of dollars worth of keywords}.
\newblock \bibinfo{journal}{\emph{American economic review}}
  \bibinfo{volume}{97}, \bibinfo{number}{1} (\bibinfo{year}{2007}),
  \bibinfo{pages}{242--259}.
\newblock


\bibitem[\protect\citeauthoryear{eMarketer}{eMarketer}{2018}]%
        {emarketer2018}
\bibfield{author}{\bibinfo{person}{eMarketer}.}
  \bibinfo{year}{2018}\natexlab{}.
\newblock \showarticletitle{US Total Media Ad Spending, by Media, 2016-2022}.
\newblock \bibinfo{journal}{\emph{eMarketer Website}} (\bibinfo{year}{2018}).
\newblock
\urldef\tempurl%
\url{https://www.emarketer.com/topics/topic/directory-ad-spending}
\showURL{%
\tempurl}


\bibitem[\protect\citeauthoryear{Getintent}{Getintent}{2017}]%
        {getintent:rtb1}
\bibfield{author}{\bibinfo{person}{Getintent}.}
  \bibinfo{year}{2017}\natexlab{}.
\newblock \showarticletitle{RTB Auctions: Fair Play?}
\newblock \bibinfo{journal}{\emph{AdExchanger}}.
\newblock
\urldef\tempurl%
\url{https://blog.getintent.com/rtb-auctions-fair-play-3b372d505089}
\showURL{%
\tempurl}


\bibitem[\protect\citeauthoryear{Google}{Google}{2019}]%
        {google:rtb}
\bibfield{author}{\bibinfo{person}{Google}.} \bibinfo{year}{2019}\natexlab{}.
\newblock \showarticletitle{Real Time Bidding Protocol, Release Notes}.
\newblock \bibinfo{journal}{\emph{Google Website}}.
\newblock
\urldef\tempurl%
\url{https://developers.google.com/authorized-buyers/rtb/relnotes#updates-2019-03-13}
\showURL{%
\tempurl}


\bibitem[\protect\citeauthoryear{Horta{\c{c}}su, Kastl, and
  Zhang}{Horta{\c{c}}su et~al\mbox{.}}{2018}]%
        {hortaccsu2018bid}
\bibfield{author}{\bibinfo{person}{Ali Horta{\c{c}}su}, \bibinfo{person}{Jakub
  Kastl}, {and} \bibinfo{person}{Allen Zhang}.}
  \bibinfo{year}{2018}\natexlab{}.
\newblock \showarticletitle{Bid shading and bidder surplus in the us treasury
  auction system}.
\newblock \bibinfo{journal}{\emph{American Economic Review}}
  \bibinfo{volume}{108}, \bibinfo{number}{1} (\bibinfo{year}{2018}),
  \bibinfo{pages}{147--69}.
\newblock


\bibitem[\protect\citeauthoryear{Juan, Zhuang, Chin, and Lin}{Juan
  et~al\mbox{.}}{2016}]%
        {juan2016field}
\bibfield{author}{\bibinfo{person}{Yuchin Juan}, \bibinfo{person}{Yong Zhuang},
  \bibinfo{person}{Wei-Sheng Chin}, {and} \bibinfo{person}{Chih-Jen Lin}.}
  \bibinfo{year}{2016}\natexlab{}.
\newblock \showarticletitle{Field-aware factorization machines for CTR
  prediction}. In \bibinfo{booktitle}{\emph{Proceedings of the 10th ACM
  Conference on Recommender Systems}}. \bibinfo{pages}{43--50}.
\newblock


\bibitem[\protect\citeauthoryear{Kitts}{Kitts}{2019}]%
        {kitts:fpa}
\bibfield{author}{\bibinfo{person}{B. Kitts}.} \bibinfo{year}{2019}\natexlab{}.
\newblock \showarticletitle{Bidder Behavior after Shifting from Second to First
  Price Auctions in Online Advertising}.
\newblock \bibinfo{journal}{\emph{unpublished}}.
\newblock
\urldef\tempurl%
\url{http://www.appliedaisystems.com/papers/FPA_Effects33.pdf}
\showURL{%
\tempurl}


\bibitem[\protect\citeauthoryear{Pan, Mao, Ruiz, Sun, and Flores}{Pan
  et~al\mbox{.}}{2019}]%
        {pan2019predicting}
\bibfield{author}{\bibinfo{person}{Junwei Pan}, \bibinfo{person}{Yizhi Mao},
  \bibinfo{person}{Alfonso~Lobos Ruiz}, \bibinfo{person}{Yu Sun}, {and}
  \bibinfo{person}{Aaron Flores}.} \bibinfo{year}{2019}\natexlab{}.
\newblock \showarticletitle{Predicting different types of conversions with
  multi-task learning in online advertising}. In
  \bibinfo{booktitle}{\emph{Proceedings of the 25th ACM SIGKDD International
  Conference on Knowledge Discovery \& Data Mining}}.
  \bibinfo{pages}{2689--2697}.
\newblock


\bibitem[\protect\citeauthoryear{Pan, Xu, Ruiz, Zhao, Pan, Sun, and Lu}{Pan
  et~al\mbox{.}}{2018}]%
        {pan2018field}
\bibfield{author}{\bibinfo{person}{Junwei Pan}, \bibinfo{person}{Jian Xu},
  \bibinfo{person}{Alfonso~Lobos Ruiz}, \bibinfo{person}{Wenliang Zhao},
  \bibinfo{person}{Shengjun Pan}, \bibinfo{person}{Yu Sun}, {and}
  \bibinfo{person}{Quan Lu}.} \bibinfo{year}{2018}\natexlab{}.
\newblock \showarticletitle{Field-weighted factorization machines for
  click-through rate prediction in display advertising}. In
  \bibinfo{booktitle}{\emph{Proceedings of the 2018 World Wide Web
  Conference}}. \bibinfo{pages}{1349--1357}.
\newblock


\bibitem[\protect\citeauthoryear{Pownall and Wolk}{Pownall and Wolk}{2013}]%
        {pownall2013bidding}
\bibfield{author}{\bibinfo{person}{Rachel~AJ Pownall} {and}
  \bibinfo{person}{Leonard Wolk}.} \bibinfo{year}{2013}\natexlab{}.
\newblock \showarticletitle{Bidding behavior and experience in internet
  auctions}.
\newblock \bibinfo{journal}{\emph{European Economic Review}}
  \bibinfo{volume}{61} (\bibinfo{year}{2013}), \bibinfo{pages}{14--27}.
\newblock


\bibitem[\protect\citeauthoryear{Rubicon}{Rubicon}{2018}]%
        {rubicon:EMR}
\bibfield{author}{\bibinfo{person}{Rubicon}.} \bibinfo{year}{2018}\natexlab{}.
\newblock \showarticletitle{Bridging the Gap to First-Price Auctions: A
  Buyer’s Guide}.
\newblock \bibinfo{journal}{\emph{Rubicon Website}}.
\newblock
\urldef\tempurl%
\url{http://go.rubiconproject.com/rs/958-XBX-033/images/Buyers_Guide_to_First_Price_Rubicon_Project.pdf}
\showURL{%
\tempurl}


\bibitem[\protect\citeauthoryear{Sluis}{Sluis}{2017}]%
        {sluis:guardiansuesrubicon}
\bibfield{author}{\bibinfo{person}{S. Sluis}.} \bibinfo{year}{2017}\natexlab{}.
\newblock \showarticletitle{Explainer: More On The Widespread Fee Practice
  Behind The Guardian's Lawsuit Vs. Rubicon Project}.
\newblock \bibinfo{journal}{\emph{AdExchanger}}.
\newblock
\urldef\tempurl%
\url{https://adexchanger.com/ad-exchange-news/explainer-widespread-fee-practice-behind-guardians-lawsuit-vs-rubicon-project/}
\showURL{%
\tempurl}


\bibitem[\protect\citeauthoryear{Tian}{Tian}{2009}]%
        {tian2009forecasting}
\bibfield{author}{\bibinfo{person}{Jing Tian}.}
  \bibinfo{year}{2009}\natexlab{}.
\newblock \showarticletitle{Forecasting the unemployment rate when the forecast
  loss function is asymmetric}.
\newblock \bibinfo{journal}{\emph{School of Economics and Finance, Faculty of
  Business, University of Tasmania}} (\bibinfo{year}{2009}).
\newblock


\bibitem[\protect\citeauthoryear{Wilkens}{Wilkens}{2017}]%
        {wilkens:cinderella}
\bibfield{author}{\bibinfo{person}{Cavallo-R. Niazadeh~R. Wilkens, C.}}
  \bibinfo{year}{2017}\natexlab{}.
\newblock \showarticletitle{GSP: The Cinderella of Mechanism Design}.
\newblock \bibinfo{journal}{\emph{Proceedings of the 26th ACM International
  Conference on World Wide Web}}, \bibinfo{pages}{25--32}.
\newblock


\bibitem[\protect\citeauthoryear{Wu, Yeh, and Chen}{Wu et~al\mbox{.}}{2018}]%
        {wu2018deep}
\bibfield{author}{\bibinfo{person}{Wush Wu}, \bibinfo{person}{Mi-Yen Yeh},
  {and} \bibinfo{person}{Ming-Syan Chen}.} \bibinfo{year}{2018}\natexlab{}.
\newblock \showarticletitle{Deep censored learning of the winning price in the
  real time bidding}. In \bibinfo{booktitle}{\emph{Proceedings of the 24th ACM
  SIGKDD International Conference on Knowledge Discovery \& Data Mining}}.
  \bibinfo{pages}{2526--2535}.
\newblock


\bibitem[\protect\citeauthoryear{Wu, Yeh, and Chen}{Wu et~al\mbox{.}}{2015}]%
        {wu2015predicting}
\bibfield{author}{\bibinfo{person}{Wush Chi-Hsuan Wu}, \bibinfo{person}{Mi-Yen
  Yeh}, {and} \bibinfo{person}{Ming-Syan Chen}.}
  \bibinfo{year}{2015}\natexlab{}.
\newblock \showarticletitle{Predicting winning price in real time bidding with
  censored data}. In \bibinfo{booktitle}{\emph{Proceedings of the 21th ACM
  SIGKDD International Conference on Knowledge Discovery and Data Mining}}.
  \bibinfo{pages}{1305--1314}.
\newblock


\bibitem[\protect\citeauthoryear{Zulehner}{Zulehner}{2009}]%
        {zulehner2009bidding}
\bibfield{author}{\bibinfo{person}{Christine Zulehner}.}
  \bibinfo{year}{2009}\natexlab{}.
\newblock \showarticletitle{Bidding behavior in sequential cattle auctions}.
\newblock \bibinfo{journal}{\emph{International Journal of Industrial
  Organization}} \bibinfo{volume}{27}, \bibinfo{number}{1}
  (\bibinfo{year}{2009}), \bibinfo{pages}{33--42}.
\newblock


\end{thebibliography}


\end{document}